\documentclass[twocolumn,showpacs,preprintnumbers,amsmath,amssymb,prd]{revtex4}

\usepackage{graphicx}
\usepackage{dcolumn}
\usepackage{bm}

\usepackage[usenames]{color}
\begin{document}

\preprint{APS/123-QED}

\title{Results of the IGEC-2 search for gravitational wave bursts\\during 2005}

\author{
P. Astone$^1$, D. Babusci$^2$, L.Baggio$^{3}$, M. Bassan$^{4,5}$, M. Bignotto$^{6,7}$, M. Bonaldi$^{8,9}$, M. Camarda$^{10}$ , P. Carelli$^{5,11}$, G. Cavallari$^{12}$,
F. Cavanna$^{11,13}$, M. Cerdonio$^{6,7}$, A. Chincarini$^{14}$, E. Coccia$^{4,13}$,  L. Conti$^{6,7}$, S. D'Antonio$^5$, M. De Rosa$^{15,16}$, M. Drago$^{6,7}$,
F. Dubath$^{17}$, V. Fafone$^{4,5}$, P. Falferi$^{8,9}$,  S. Foffa$^{17}$, P. Fortini$^{18}$, S. Frasca$^{1,19}$, G. Gemme$^{14}$, G. Giordano$^2$, G. Giusfredi$^{20}$,
W.O. Hamilton$^{21}$,  J. Hanson$^{21}$, M. Inguscio$^{16,22}$, W.W. Johnson$^{21}$, N. Liguori$^{6,7}$, S. Longo$^{23}$,
M. Maggiore$^{17}$, F. Marin$^{16,22}$, A. Marini$^2$, M. P. McHugh$^{24}$, R. Mezzena$^{9,25}$,  P. Miller$^{21}$,  Y. Minenkov$^{13}$, A. Mion$^{9,25}$,
 G. Modestino$^2$, A. Moleti$^{4,5}$, D. Nettles$^{21}$, A. Ortolan$^{23}$, O. Palamara$^{13}$,
G.V. Pallottino$^{1,19}$, R. Parodi$^{14}$, G. Piano Mortari$^{11,13}$, S. Poggi$^{26}$, G.A. Prodi$^{\ast 9,25}$, L. Quintieri$^2$, V. Re$^{9,25}$,
A. Rocchi$^{4}$, F. Ronga$^2$, F. Salemi$^{9,25}$, G. Soranzo$^{7}$, R. Sturani$^{17}$, L. Taffarello$^{7}$, R. Terenzi$^{1,27}$, G. Torrioli$^{1,28}$, R. Vaccarone$^{14}$,
G. Vandoni$^{12}$, G. Vedovato$^{7}$, A. Vinante$^{8,9}$, M. Visco$^{4,27}$, S. Vitale$^{9,25}$, J. Weaver$^{21}$, J.P. Zendri$^{7}$ and P. Zhang$^{21}$ }

\address{(IGEC-2 Collaboration)}

\address{$^1$ INFN, Sezione di Roma, P.le A.Moro 2, I-00185, Roma, Italy}
\address{$^2$ INFN, Laboratori Nazionali di Frascati, Via E.Fermi 40, I-00044, Frascati, Italy}
\address{$^{3}$ EGO, 56021 S. Stefano a Macerata, Cascina, Pisa, Italy}
\address{$^4$ Dipartimento di Fisica, Universit\`a di Roma ``Tor Vergata'', Via Ricerca Scientifica 1,  I-00133 Roma, Italy}
\address{$^5$ INFN, Sezione di Roma Tor Vergata, Via Ricerca Scientifica 1, I-00133 Roma, Italy}
\address{$^6$ Dipartimento di Fisica, Universit\`a di Padova, Via Marzolo 8, 35131 Padova, Italy}
\address{$^7$ INFN, Sezione di Padova, Via Marzolo 8, 35131 Padova, Italy}
\address{$^8$ Istituto di Fotonica e Nanotecnologie, CNR-Istituto Trentino di Cultura, I-38050 Povo (Trento), Italy}
\address{$^9$ INFN, Gruppo Collegato di Trento, Sezione di Padova, I-38050 Povo, Trento, Italy}
\address{$^{10}$ Dipartimento di Ingegneria Informatica, Universit\`a di Padova, Via G. Gradenigo 6a, 35131 Padova, Italy}
\address{$^{11}$ Dipartimento di Fisica, Universit\`a de L'Aquila, L'Aquila, Italy}
\address{$^{12}$ CERN, Geneva, Switzerland}
\address{$^{13}$ INFN, Laboratori Nazionali del Gran Sasso, Assergi, L'Aquila,Italy}
\address{$^{14}$ INFN, Sezione di Genova, Genova, Italy}
\address{$^{15}$ INOA, CNR, I-80078 Pozzuoli, Napoli, Italy}
\address{$^{16}$ INFN, Sezione di Firenze, I-50121 Firenze, Italy}
\address{$^{17}$ D$\acute{e}$partement de Physique Th$\acute{e}$orique, Universit$\acute{e}$ de Gen\`eve, Gen\`eve, Switzerland }
\address{$^{18}$ Dipartimento di Fisica, Universit\`a di Ferrara and INFN, Sezione di Ferrara, I-44100 Ferrara, Italy}
\address{$^{19}$ Dipartimento di Fisica, Universit\`a di Roma ``La Sapienza'', P.le A.Moro 2, I-00185, Roma, Italy}
\address{$^{20}$ INOA, CNR, I-50125 Arcetri, Firenze, Italy}
\address{$^{21}$ Department of Physics and Astronomy, Louisiana State University, Baton Rouge, Louisiana 70803}
\address{$^{22}$ LENS and Dipartimento di Fisica, Universit\`a di Firenze, I-50121 Firenze, Italy}
\address{$^{23}$ INFN, Laboratori Nazionali di Legnaro, 35020 Legnaro, Padova, Italy}
\address{$^{24}$ Department of Physics, Loyola University New Orleans, New Orleans, LA 70118,USA}
\address{$^{25}$ Dipartimento di Fisica, Universit\`a di Trento, I-38050 Povo, Trento, Italy}
\address{$^{26}$ Consorzio Criospazio Ricerche, I-38050 Povo, Trento, Italy}
\address{$^{27}$ INAF, Istituto Fisica Spazio Interplanetario,Via Fosso del Cavaliere, I-00133 Roma, Italy}
\address{$^{28}$ CNR, Istituto di Fotonica e Nanotecnologie,  Roma, Italy}

\email[Corresponding author ]{prodi@science.unitn.it}


\begin{abstract}
The network of resonant bar detectors of gravitational waves resumed coordinated observations within the International Gravitational Event Collaboration (IGEC-2). Four detectors are taking part in this collaboration: ALLEGRO, AURIGA, EXPLORER and NAUTILUS. We present here the results of the search for gravitational wave bursts over 6 months during 2005, when IGEC-2 was the only gravitational wave observatory in operation. The network data analysis implemented is based on a time coincidence search among AURIGA, EXPLORER and NAUTILUS, keeping the data from ALLEGRO for follow-up studies. With respect to the previous IGEC 1997-2000 observations, the amplitude sensitivity of the detectors to bursts improved by a factor $\approx 3$ and the sensitivity bandwidths are wider, so that the data analysis was tuned considering a larger class of detectable waveforms. Thanks to the higher duty cycles of the single detectors, we decided to focus the analysis on three-fold observation, so to ensure the identification of any single candidate of gravitational waves (gw) with high statistical confidence. The achieved false detection rate is as low as 1 per century. No candidates were found.

\end{abstract}

\maketitle

\section{\label{sec:intro}Introduction}
The search for transient gravitational waves (gws) requires the use of a network of detectors. In fact, the analysis of simultaneous data from more detectors at different sites allows an efficient rejection of the spurious candidates, either caused by transient local disturbances or by the intrinsic noise of the detectors. Moreover, the false alarm probability of the network due to uncorrelated noise sources at the different sites can be reliably estimated. 

The first long term search for bursts gw by a network of detectors has been performed by the five resonant bars ALLEGRO, AURIGA, EXPLORER, NAUTILUS and NIOBE, within the International Gravitational Event Collaboration (IGEC)~\cite{IGECPRL}. The search consisted in a time coincidence analysis over a 4-year period, from 1997 to 2000, and set an upper limit on the rate of impulsive gravitational waves as a function of the gw amplitude threshold of the data analysis~\cite{IGEC}. However, the overlap in observation time of the detectors was modest: three or more detectors were in simultaneous validated observation for $173~days$, $\simeq12\%$ of the time, and two-fold observations covered an additional period of $534~days$, $\simeq36\%$. Moreover, since in the two-fold coincidence searches at the lowest amplitude thresholds some false alarms were expected, the IGEC 1997-2000 observation was not able to discriminate a single gw candidate from the accidental coincidences for most of the time. The target gw signals were short transients showing a flat Fourier component around 900 Hz, like pulses of $\sim 1~ms$ duration or oscillating signals with a few cycles of $\sim 1~ms$ period.

The same class of signals was targeted in the searches
performed with the EXPLORER and NAUTILUS data in 2001~\cite{ROGrun2001}
and 2003~\cite{ROGrun2003}. These searches, being based on
two-fold coincidences, could not aim at the identification
of single GW candidates. They addressed the study of a
possible excess of coincidences taking advantage of sidereal time analysis.

Subsequent searches for burst gws have been performed also by networks of interferometric detectors, which feature a better sensitivity in a wider frequency bandwidth. In particular, the LIGO searches demonstrated a significant improvement in sensitivity from the 2003 observation~\cite{LIGOs2} to the 2005 observation~\cite{LIGOs4}.
Due to the shorter duration of these searches, an improvement on the limit set by IGEC at higher amplitudes on the rate of millisecond gw signals was not possible. However, In November 2005, the LIGO observatory started its first long term scientific observation at its design sensitivity~\cite{LIGOShh}.

In 2004 four bar detectors resumed simultaneous operation: ALLEGRO~\cite{ALLEGRO}, AURIGA~\cite{status-AURIGA-2003,Amaldi-Andrea}, EXPLORER and NAUTILUS~\cite{ROGa,ROGb}. A new long term gw search started under the IGEC-2 Collaboration, whose primary goal is to identify any single candidate of burst gw with high statistical confidence. This coordinated observation is still running and targets to a broader signal class than the previous IGEC search, as for instance binary BH mergers and ring-downs~\cite{BHcampanelli} and longer transients recently predicted for Supernova core collapses~\cite{OttBurrows06}.

This paper is the first report on the IGEC-2 observations and describes the results of the analysis of 6 months of data, from May the $20^{th}$ to November the $15^{th}$ 2005, when IGEC-2 was the only gravitational wave observatory in operation. The AURIGA, EXPLORER and NAUTILUS data are actually used to search for gravitational wave candidates showing up as triple time coincidences. Due to a delay in the validation of the ALLEGRO data, we agreed to use the data from this detector for follow-up investigations on possible signal candidates identified by the other resonant bars. 

\section{\label{sec:goal}Characteristics and goals of the IGEC-2 observatory}
\begin{figure}
\includegraphics[width=20pc]{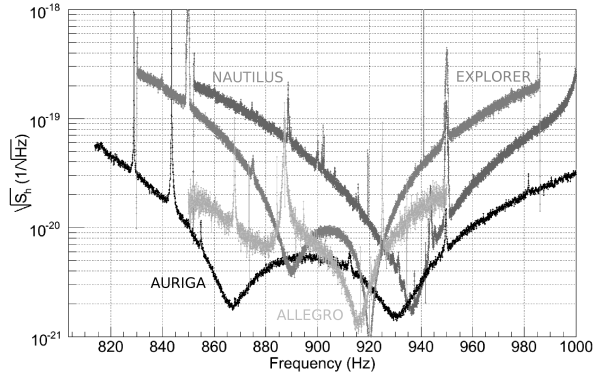}
\caption{\label{fig:Shh} Typical strain noise spectral density (single-sided) curves of IGEC-2 detectors in 2005. From light gray to black: ALLEGRO, EXPLORER, NAUTILUS and AURIGA. All detectors share comparable minimum levels of noise spectra. The wider bandwidth of AURIGA includes the bandwidths of the other detectors.}
\end{figure}

As for the previous IGEC search in 1997-2000, the detectors are aligned within a few degrees and so feature the same directional sensitivity at any time. The spectral sensitivities of the resonant bar detectors during 2005 is shown in Fig.~\ref{fig:Shh}. 
 The minima of the noise power spectral densities are very close, within $1\div2\times 10^{-21} 1/\sqrt{Hz}$, as  the four detectors share a similar design (i.e. cylindrical Al-5056 bar with a mass of $\simeq 2200 ~kg$ cooled at liquid He temperature, resonant transducers, similar mechanical quality factors, dc-SQUID signal amplifier).
 With respect to the previous IGEC-1 network ~\cite{IGEC}, all detectors have been upgraded and exhibit now wider bandwidths. EXPLORER  and NAUTILUS were improved respectively in 2000 and 2002. Some modifications on the cryogenics apparatus and mechanical filters, new  transducers and dc-SQUIDs were adopted~\cite{ROG-PRL}.  The upgrade of AURIGA, completed in 2003, concerned most of the apparatus, from seismic isolation system~\cite{AUsospensioni} to the readout~\cite{AUsquid}; in particular, a better coupling between the transducer and the signal amplifier was achieved by tuning the electric resonance of the signal transformer to the mechanical modes of the antenna and transducer and the signal amplifier is now based on a two stage dc-SQUID. The result was a very large increase in the detector bandwidth~\cite{3-modes,Amaldi-Andrea}. Additional upgrades of the room temperature suspensions during 2005 led to a significant improvement of duty-cycle and data quality. ALLEGRO resumed operation in early 2004, after changing both the resonant transducer 
and the readout electronics ~\cite{ALLEGRO}.




The primary scientific interest of the IGEC-2 observations is the ability of identifying any single candidate of gravitational wave signal with high statistical confidence. Moreover, the results hereby presented refer to a period when only IGEC2 was surveying gws. 


\section{\label{sec:data}Characteristics of the exchanged data}
During the 180 days considered in this analysis, the AURIGA, EXPLORER and NAUTILUS detectors show a high duty cycle, see Tab.~\ref{tab:duty}. In particular the validated data of AURIGA, EXPLORER and NAUTILUS overlap by 130.71 days in three-fold coincidence (corresponding to 73\%) and 45 days more are covered as two fold coincidences (about 25\%).

\begin{figure}[ht]
\includegraphics[width=20pc]{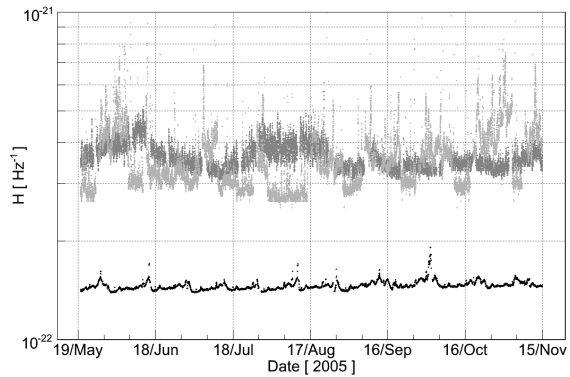}
\caption{\label{fig:sigma} Noise vs time of AURIGA (black), EXPLORER (light gray) and NAUTILUS (dark gray) detectors. The ordinate shows $1\sigma$ noise in terms of equivalent Fourier component $H$ of the strain waveform $h(t)$ of a millisecond gw pulse.}
\end{figure}

The noise stability of the detectors is remarkable, either if compared to past performances or to data analysis requirements. As shown in Fig.~\ref{fig:sigma}, the standard deviation of AURIGA noise shows a slow systematic dependence on the liquid He levels in the cryostats with peak-to-peak variations  of the order of 10 \%. Minimum noise levels of EXPLORER and NAUTILUS are higher by a factor of $\sim2$ in terms of equivalent amplitude of a millisecond gravitational wave burst.

Each group validates its data and tunes its searches for gravitational wave candidates independently. These analyses are based on linear filters matched to $\delta -like$ signals. The algorithms implemented for the AURIGA filter and for the EXPLORER and NAUTILUS filter are different and have been independently developed. In both pipelines, the filtered data stream is calibrated to give the reconstructed Fourier component $H$ of the strain waveform $h(t)$ of a short ($\delta-like)$ gravitational wave burst at input. 

A candidate event is identified by detecting a local maximum in the absolute value of the filtered data stream: the occurrence time of the maximum and the corresponding amplitude are the estimates of the arrival time and of the Fourier component $H$ of the gw $h(t)$. Since these estimates refer to the $\delta-filter$, they are consistent only for gws of short duration with a flat Fourier transform over the detection bandwidth. For waveforms of colored spectral structure, the filter mismatch leads to non-optimal SNRs (Signal to Noise Ratios) for the candidate events and to biases in their amplitude and time estimates.

As an example, in the case of signals shaped as damped sinusoids with damping time $\tau$, the typical $SNR$ reconstructed by $\delta$ filters is $\gtrsim 80\%$ of the $SNR$ of the signal-matched filter for $\tau$ $\lesssim 10 ~ms$, $\lesssim 25~ms$ and $\lesssim 50 ~ms$ for AURIGA, EXPLORER and NAUTILUS respectively. For such classes of waveforms, the implemented $\delta-filters$ reconstruct the arrival times with relative systematic errors $\lesssim \tau/2$.

A cross validation has been performed on the different analysis pipelines. A sample day of raw data of EXPLORER and of NAUTILUS was processed by AURIGA data analysis, using the same epoch vetoes, but with a different implementation of the data validation and conditioning. The comparison of the candidate events found by AURIGA and ROG pipelines show a good consistency for $SNR \geq 5 $ with some unavoidable differences at lower $SNR$.

\begin{table}[!htbp]
\caption{\label{tab:duty} Overview of the validated observation periods for the 180 days considered in this analysis. Off-diagonal terms show the two-fold coincidence times. }
\begin{ruledtabular}
\begin{tabular}{|c|c|c|c|}
 &AURIGA & EXPLORER & NAUTILUS \\
\hline
AURIGA & 172.9 d &  &  \\
EXPLORER & 151.8 d & 158.0 d &  \\
NAUTILUS & 150.2 d & 135.3 d & 155.0 d\\
\end{tabular}
\end{ruledtabular}
\end{table}

Event lists of each detector are exchanged according to the protocol of the previous IGEC 1997-2000 observations~\cite{IGEC-Protocol}, and include information on amplitude and time uncertainties and on the amplitude threshold used to select the events. 

A rigid time offset, chosen within $\pm 10~s$, is added to all the event lists prior to the exchange and is kept confidential, so that all the tuning of the analysis is performed without knowledge of the true coincidences. When the network analysis is completely defined, these confidential time shifts are disclosed to draw the final results. This procedure, referred to as {\it blind} analysis, ensures an un-biased statistical interpretation of the results. 


The choice of the most suitable exchange threshold is left to each group. 
The thresholds used to select the exchanged events are $SNR=4.5$ for AURIGA and $SNR=4.0$ for EXPLORER and NAUTILUS. 
These are considered as the minimal thresholds that allow the identification of a candidate event by each detector with reasonable confidence, according to the results of tests carried out with hardware and software injections.  
In particular, at lower $SNR$ the timing uncertainty related to the candidate events increases rapidly. 

For the AURIGA events, the conservative estimates of the timing uncertainty  ($1 \sigma$) range from a maximum of $5~ms$ at the threshold to a minimum of $\sim 0.5 ~ms$ at $SNR>10$, as computed assuming $\delta -like$ signals. For EXPLORER and NAUTILUS, the $1 \sigma$ timing uncertainty was conservatively set to $10~ms$.

\begin{figure}
\includegraphics[width=20pc]{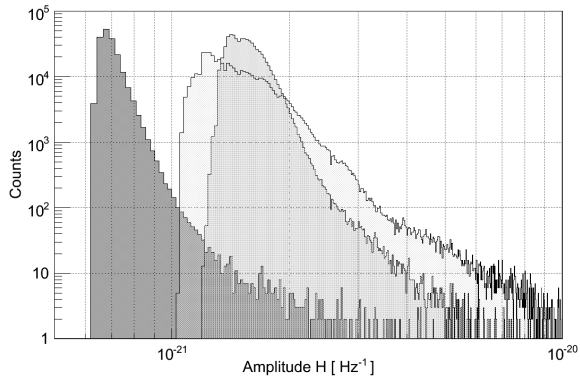}
\caption{\label{fig:histoevents} Amplitude distribution of the exchanged candidate events above the minimal thresholds: AURIGA (darker gray) $SNR > 4.5$, EXPLORER (lighter gray) $SNR > 4.0$ and NAUTILUS (gray) $SNR > 4.0$. The amplitude is given in terms of the Fourier component $H$ of the $h(t)$ waveform of a millisecond gw pulse. 
}
\end{figure}

The amplitude distribution of the exchanged events corresponding to the period of three-fold observations is shown in Fig.~\ref{fig:histoevents}. The amplitude distribution of the AURIGA events is very close to that expected for Gaussian noise up to $SNR\simeq 5.5$, while non Gaussian outliers are dominating at at higher $SNR$s. The number of candidate events above the minimal thresholds is listed in Tab.~\ref{tab:eventnum}; the mean event rate is $\sim 45/h$ for AURIGA, while it is larger for EXPLORER and NAUTILUS, $\sim 129/h$ and $\sim 183/h$ respectively, due to the lower $SNR$ thresholds.

\begin{table}
\caption{\label{tab:eventnum}Number of candidate events per each detector for some data selections considered in the network analysis. The leftmost column refers to the minimal thresholds, the central column to the event selection optimized for signals with comparable SNR, the rightmost column to the selection optimized for signals with comparable $H$ amplitudes, see Sec.~\ref{sec:selection}.}
\begin{ruledtabular}
\begin{tabular}{rrrr}
data cut AU & $SNR>4.5$ & $SNR>4.95$ & $SNR>7.0$  \\
 EX & $SNR>4.0$ & $SNR>4.95$ & $SNR>4.25$  \\
 NA & $SNR>4.0$ & $SNR>4.95$ & $SNR>4.25$  \\
\hline
AURIGA & 186911 & 34598 & 790 \\
EXPLORER & 489103 & 29217 & 245000 \\
NAUTILUS & 679775 & 42028 & 351375 \\
\end{tabular}
\end{ruledtabular}
\end{table}

\section{\label{sec:analysis}Network data analysis}
The network data analysis consists of a time coincidence search among 
the exchanged events. The coincidence time window is set accordingly to
the same procedure previously implemented within IGEC 1997-2000 search ~\cite{IGEC}. Two events are defined in coincidence if their arrival times $t_i$ and $t_j$ are compatible within their variances, $\sigma _i ^2$ and $\sigma _j ^2$ : 

\begin{equation}
|t_i - t_j|  < k \ \sqrt{\sigma _i ^2 + \sigma _j ^2}.
\label{eq:coinc}
\end{equation}
 where $k$ is set to $4.47$, as in ref.~\cite{IGEC}. According to the Byenaim\'e-Tchebychev inequality (see for example ~\cite{Papoulis}), this choice limits the maximum false dismissal probability of the above coincidence condition to 5\% regardless of the statistical distribution of arrival time uncertainties. For the three-fold coincidence search considered here, the same condition is required per each detector pair, leading to a maximum false dismissal probability $< 1-0.95^{~3} \sim 14\%$. The resulting coincidence windows are $\simeq 63 ~ms$ between EXPLORER and NAUTILUS and $45 - 50 ~ms$ when AURIGA is considered. In the previous searches performed between Explorer and Nautilus
a fixed time window of 30 ms was adopted by the ROG group.
This value ensured a low false dismissal in the case of
delta-like signals considering the measured time response to
excitations due to cosmic ray showers \cite{ROGa}.

Here we have neglected the effect of the propagation time of the gws among the different sites, since it is quite small, $\leq 2.4 ~ms$. Moreover, in case the signal duration is not small with respect to the coincidence window, the cited false dismissals are no more strictly ensured, because the systematic uncertainties on the arrival time can be different in different detectors (see the previous section).


 The coincidence search is tuned to ensure a high statistical 
confidence in case of detection of any single gravitational wave: 1 false 
alarm per century. To meet this requirement, we analyze only the 130.71 
days of three-fold observation by AURIGA, EXPLORER and NAUTILUS, since 
the two-fold coincidence search cannot reach such a low rate of accidental 
coincidences without sacrificing too much on the sensitivity side. 


\subsection{\label{sec:accidental}Accidental coincidence estimates}

\begin{figure}
\includegraphics[width=20pc]{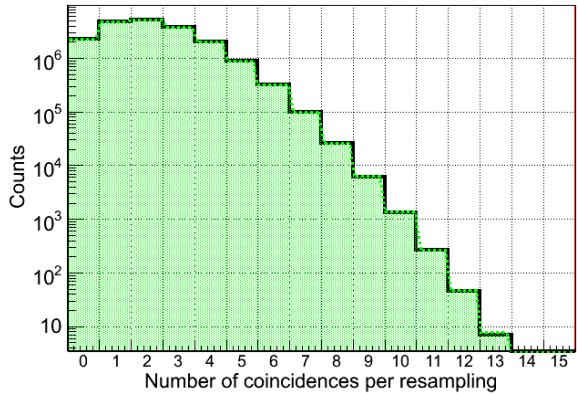}
\caption{\label{fig:histoacci} Histogram of the number of accidental coincidences per each resampling (black continuous line). Exchanged events with $SNR>4.0$ for EXPLORER and NAUTILUS and $SNR>4.5$ for AURIGA have been considered. 19'355'600 independent off-source resampling of the experiment have been computed by shifting the time of two detectors within $\pm11000~s$ in $5 s$ steps, excluding the central region around the nominal zero-lag (see Sec.~\ref{sec:data}). The histogram is well in agreement with the Poisson distribution with mean equal to 2.16 (gray dotted line and shaded area), as $\chi ^2 = 11.3$ with 12 degrees of freedom corresponding to a p-value of $50.3\%$.}
\end{figure}

The three-fold accidental coincidences has been investigated with high statistics: about 20 millions independent, {\it off-source}, resampling of the counting experiment have been performed by applying relative shifts at the times of two detectors within
$\pm11000 s$ in $5 s$ steps. The resulting changes in the overlap
time of the resamplings with respect to the actual observation time are
negligible: the mean observation time of the resamplings is $\sim 0.09\%$ less than the actual observation time and the largest difference is at most $\sim 0.4\%$.
Fig.~\ref{fig:histoacci} shows the histogram of the counts of the accidental coincidences using the whole set of the exchanged events. The histogram is very well in agreement with a Poisson distribution of mean equal to 2.16 counts per observation time. 

Cross-checks on the accidental coincidences rate estimate have been
performed with other independent algorithms and different choices of the relative shifts, giving in all cases
results well within the expected statistical fluctuations. 
A further check was pursued with a method based on an analytic estimate
of the random coincidence rates (see Appendix). The
results were in very good agreement with the values obtained
with the time-shifts technique.

\subsection{\label{sec:selection}Data selection}
In order to achieve the goal of 1 false alarm per century, a data selection is necessary to reduce by a factor $\approx 600$ the accidental coincidences found on the exchanged data set. In general, the data selection has to be tuned with the aim of preserving the detection efficiency of the gw survey. In our case the balance between false alarms and detection efficiency has been addressed from first principles, since measurements of average efficiency during the observation time were not available. We decided to perform three searches based on different data selection procedures:

\begin{itemize}
\item A) fixed and equal thresholds on the $SNR$ of the events of each detector. Its motivation relies on setting a minimal comparable significance for the considered events as well as setting a similar events rate for each detector. Given the different spectral sensitivities, this search is more sensitive to colored signals that show the largest fraction of their power in the overlapping part of the bandwidths (e.g. $915-945 Hz$, see Fig.~\ref{fig:Shh}) rather than in the remaining part of the AURIGA bandwidth. Such signals would produce similar $SNRs$ in the $\delta$ filtered data of all detectors.
\item B) fixed thresholds on the $SNR$ of the events, but chosing different $SNR$ thresholds for the detectors so that they correspond to comparable levels of absolute gw amplitude $H$ in all detectors. This search is targeted to short signals which feature a flat Fourier transform within the AURIGA bandwidth and therefore appear at higher $SNR$ in AURIGA with respect to EXPLORER and NAUTILUS. It allows to use lower $SNR$ thresholds for EXPLORER and NAUTILUS than the previous data selection procedure.
\item C) common absolute amplitude thresholds: same procedures used in the IGEC 1997-2000 search. The different data sets are selected according to a common gw amplitude $H_i$~\cite{IGEC}: the coincidence search is performed only during the periods when the exchange thresholds of all detectors were lower than $H_i$ and considering only the events whose amplitude is larger than $H_i$. This procedure is repeated for a grid of selected $H_i$ values. This search, as the previous one, is targeted to short bursts. Differently from the two previous procedures, this search not only selects the events, but also the effective observation time as a function of $H_i$. Its main advantages are to keep under control the false dismissal probability of the observatory and therefore to make possible an interpretation in terms of rate of gw candidates as well as a straightforward comparison with the previous IGEC upper limit results.
\end{itemize}

We decided to consider the union of these three searches, i.e. to perform one composite search made by an "OR" of the three data selections procedures. This new approach simplifies the statistical analysis, since it takes care of the correlations expected in our multiple trials. In fact, any accidental coincidence occurring in more trials is counted only once and the expected overall distribution of accidental coincidences is estimated by histogramming the union of the found accidental coincidences on off-source samples. 

Our tuning led to the following choices of data selections: A) $SNR>4.95$; B) AURIGA $SNR>7.0$, EXPLORER and NAUTILUS $SNR>4.25$, C) common search thresholds $H_i = 1.2, \ 1.3, \ 1.4, \ ..., \ 3.0 \times10^{-21} /Hz$. Tab.~\ref{tab:eventnum} reports the number of considered events of each detector for data selections A) and B). In particular in A) the event rate is similar in all detectors even though AURIGA features a wider bandwidth. Instead in B), the number of AURIGA events is a few hundred times smaller that that of EXPLORER and NAUTILUS.

There is a significant correlation in false alarms between data selections B) and C), which show a large fraction of common accidental coincidences. Tab.~\ref{tab:fanum} summarizes the numbers of accidental coincidences found for the three data selections on the same off-source resampling considered in Sec.~\ref{sec:accidental}. 

\begin{table}
\caption{\label{tab:fanum} Number of accidental coincidences found on the 19355600 off-source resamplings per each data selection procedure (diagonal). The accidental coincidences found in common between different data selections are reported off-diagonal. Data selections B and C feature an evident correlation of their accidental noises. The false alarm rates of each trial taken separately are 0.396, 0.573 and 0.134 per century for A, B and C respectively. The resulting false alarm rate of the composite search, $A \cup B \cup C$ is $1.01$ per century.}
\begin{ruledtabular}
\begin{tabular}{rrrr}
 AU & $SNR>4.95$ & $SNR>7.0$ & common\\
 EX & $SNR>4.95$ & $SNR>4.25$ & search \\
 NA & $SNR>4.95$ & $SNR>4.25$ &  threshold \\ 
 data cut & A & B & C \\
 \hline
A & 27368 &   &   \\
B & 515 & 39507 &  \\
C & 147 & 5177 & 9280 \\
\end{tabular}
\end{ruledtabular}
\end{table}

\begin{figure}
\includegraphics[width=20pc]{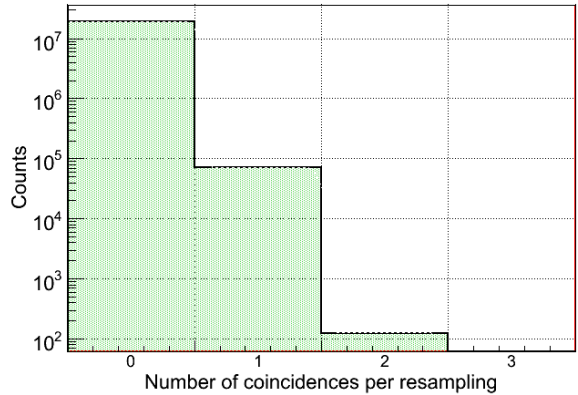}
\caption{\label{fig:histofa} Histogram of the number of accidental coincidences found per each off-source resampling for the composite search made by the union of three data selection procedures (see Sec.~\ref{sec:selection}). The histogram (black continuous line) is in agreement with a Poisson distribution with mean 0.00364 (gray dotted line and shaded area, $\chi^2=0.06$ with 1 degree of freedom), which is taken as the reference distribution for the coincidences assuming that only accidental coincidences are present. The false detection probability is to 0.00363, corresponding to 1.01 false alarms per century.}
\end{figure}

The final histogram of the accidental coincidences is plotted in Fig.~\ref{fig:histofa}: the probability of getting a non zero number of accidental coincidences is 0.00363 during the observation time, corresponding to 1.01 false alarms per century. The estimated $1\sigma$ statistical uncertainty on this probability is $2\times10^{-5}$. This uncertainty has been empirically determined by grouping the off-source samples in many disjoint subsets of equal size. The standard deviation of the number of accidental coincidences in these subsets  has been propagated to the mean, calculated on the entire off-source dataset. The resulting $\sigma$ is only slightly higher, by a factor $\simeq 1.4$, than the one expected from a purely Poisson model. Independent checks with different pipelines and on different sets of off-source samples limit the systematic uncertainty on the probability to $\lesssim 1\times10^{-4}$.

\subsection{\label{sec:plan}Plan of the statistical data analysis}

Before looking at the true coincidences in the \textit{on-source} data set, we finalize a priori the plan for the statistical data analysis. Two steps are planned: the test of the null hypothesis and the setting of confidence intervals.

We chose to reject the null if at least one triple coincidence is found in the on-source data set of the composite search. This corresponds to a significance of the test of $99.637\%$ with a $3\sigma$ statistical uncertainty of $\pm0.006\%$. Therefore, if at least one coincidence is found, the collaboration excludes it is an accidental coincidence with the above stated confidence: 
In fact, the rejection of the null points out a correlation in the observatory at the true time (i.e. not consistent with the measured accidental noise at different time lags). The source of correlation may be gws or disturbances affecting distant detectors (e.g. instrumental correlations).



The final result on the estimated number of coincidences, related to any source of correlated noise or gws, is given by confidence intervals ensuring a minimum coverage, i.e. the probability that the true value is included in that interval. We decided to set confidence intervals according to the standard confidence belt construction of Feldman and Cousins \cite{Feldman-Cousins}. The noise model for the number of coincidences is the Poisson distribution shown in Fig.~\ref{fig:histofa}. To take into account its uncertainties, we consider the union of the confidence belts given by the mean noise $\pm 3 \sigma$, i.e.~$0.00364\pm0.00006 \ events$. Thanks to this low false alarm rate, the chosen confidence belt detaches from 0 when at least one coincidence is found (provided that the coverage is lower than the significance of the null hypothesis test). The final result cannot be easily interpreted in terms of gw source models, since IGEC-2 is lacking a measurement of the detection efficiency.

Any triple coincidence found would then be investigated \textit{a posteriori} using also the data set of ALLEGRO, whenever possible. These follow-up results would be interpreted in terms of likelihood or subjective confidence by the collaboration and would not affect the significance of the rejection of the null hypthesis. The main goal of the follow-up investigation will be to discriminate among known possible sources, e.g. gravitational waves,  electromagnetic or seismic disturbances, etc.. An exchange of the raw data and gw transfer function would allow to implement more advanced network analyses, as searches based on cross-correlation.  Additional complementary information could come from electromagnetic and neutrino detectors as well as from environmental monitors.

\section{\label{sec:results}Results}

Once the network analysis was tuned, the groups exchanged the confidential time shifts necessary to reconstruct the on-source data set. This blind procedure makes the statistical interpretation of any result unambiguous.

No triple coincidences are found in the composite search described in the previous Sections and therefore the null hypothesis is not rejected. 

The upper limit set by the full search is given in terms of the number of detectable gravitational wave candidates, since the false dismissal of the composite search has not been directly measured for any model of gw source. According to the chosen confidence belt, the upper limits are $\simeq 2.4$ and $3.1$ events at 90\% and 95\% coverages respectively. For a gw waveform with a flat Fourier transform over the bars bandwidths, the efficiency of this search is mainly contributed by the data selection B (see sec.~\ref{sec:selection}). In this case, according to back of the envelope calculations, IGEC-2 features a low false dismissal, $\lesssim 0.1$, at Fourier amplitudes $\gtrsim 2\times 10^{-21} Hz^{-1}$ for optimally oriented sources. 


Outside the planned composite search for gws, we checked a posteriori  the number of coincidences among all exchanged events.  Three coincidences were found, well in agreement with the expected Poisson distribution of mean 2.16, as presented in subsection~\ref{sec:accidental}. All the events associated with these three-fold coincidences were at $SNR$ close to the thresholds and therefore no follow-up investigation has been implemented for diagnostic purposes.

\subsection{\label{subsec:comparison}Comparison with IGEC previous results}
\begin{figure}
\includegraphics[width=24pc]{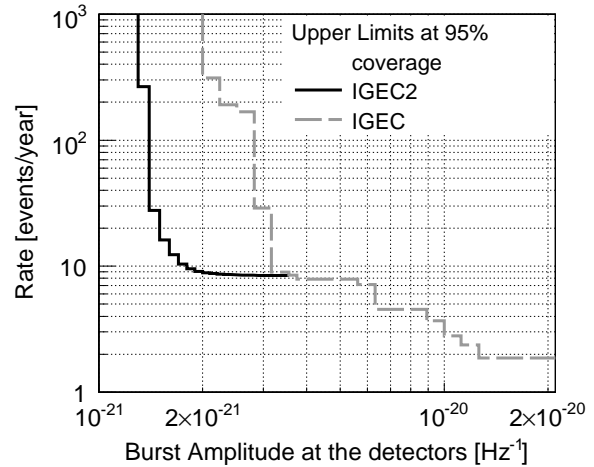}
\caption{\label{fig:ul} Comparison of the current upper limit with the previous one set on 1997-2000 observations. The uninterpreted upper limits have been computed with the same methodology. The new result however uses only a subset (i.e. the C data selection, see Sec.~\ref{sec:data})of the composite search performed on 2005 data. }
\end{figure}

Using the subset of the current results relative to the data cut C, we can compute the upper limit on the rate of millisecond bursts as a function of the amplitude threshold. This upper limit is uninterpreted, i.e. it is set in terms of detectable gws, and is done mainly for comparison with the previous IGEC 1997-2000 search~\cite{IGEC}, see Fig.~\ref{fig:ul}. The new upper limit improves the old one at lower amplitudes thanks to the better sensitivity of current detectors. The current asymptotic rate, $\simeq8.4 events/yr$ at 95\% coverage, is higher than in the previous search because of the shorter observation time, but it is reached at much lower signal amplitudes. In fact, the current detectors feature much more stationary performances and the current search is free from false alarms, while the 1997-2000 result was dominated at low amplitudes by two-fold observations, which typically show several false alarms per year.

As a general remark, the main improvement of the current result is the capability of identification of any single candidate gw, while the previous upper limit was mostly contributed by coincidence searches with much higher false alarm rates. An additional improvement of the current search comes from the new data selections procedures (i.e. data selections A and B), which extend the target towards a broader class of signals. 

\section{\label{sec:remarks}Final Remarks}
The IGEC-2 observatory is currently surveying for gw transients. Our results show that for a plain time coincidence search at least three-fold bar observations are necessary to identify any single candidate gw with satisfactory statistical confidence. The role of the resonant bar observatory is significative to search for signals occurring whenever the network of the more sensitive interferometric detectors is not fully operative and therefore not able to issue an autonomous detection of a gw candidate. In fact, since the spectral sensitivity achieved by the LIGO instruments is better than a factor $\sim10$ at the narrower bandwidths of the IGEC2 detectors, LIGO is nowadays able to perform surveys and, eventually, set upper limits at lower amplitudes and on a wider class of gws signals than IGEC-2~\cite{LIGOs4}.  In this framework, IGEC-2 can collaborate with the other observatories to extend the time coverage of current gw surveys and can contribute to the identification of rare gw events. 
In addition, if a candidate gw will be identified by the interferometric observatory, a joint investigation bar-interferometer could increase the information on the gw candidate, for instance on the signal direction and polarization amplitudes. To take the most from an hybrid bar-interferometer observatory, the data analysis methodology should overcome the intrinsic limitations of a time coincidence search and exploit the phase information of the $h(t)$ data streams provided by the different detectors, aiming at the solution of the inverse problem for the wave tensor. Tests of such methodologies are ongoing using short periods of real data sets.

\appendix
\section{analytical estimate of the accidental coincidences}
The rate of accidental coincidences of the IGEC-2 observatory has been empirically estimated by shifting the time of the detectors' data. The results have been checked by comparison with the following analytical method, based on the common assumption that the exchanged events are Poisson point processes with a slowly variable rate. 

\subsection{Analytical model}
The expected number of accidental coincidences $N_{acc}$ in the simpler case of a constant time coincidence window $\pm w$ and constant event rate is
\begin{equation}
N_{acc} = M \left( \frac{w}{T_{obs}} \right)^{M-1} \prod_{i=1}^{M} N_i
\label{eq_3std}
\end{equation}
where $M$ is the number of detectors, $T_{obs}$ is the
common observation time, and $N_i$ the number of events
in the $i^{th}$ detector.


In our case, the coincidence window $w$ depends on the detector pair $i,j$ and changes for the different AURIGA events (see eq.1 and sec.\ref{sec:data}). Therefore, the above expression for $N_{acc}$ has to be modified as shown in the following.

\begin{figure}
\vspace*{-10mm}
\includegraphics[width=20pc]{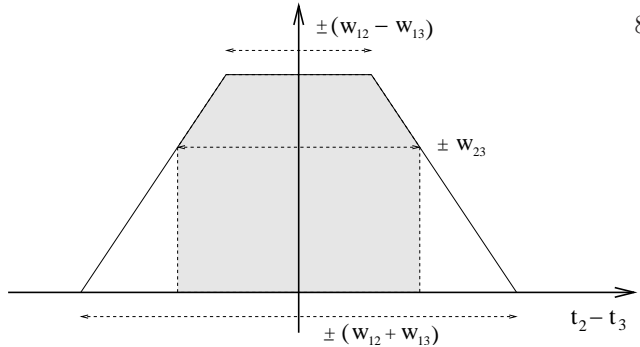}
\caption{P.d.f. of the variable $x=t_2-t_3$. The area of the
part of the trapezium inside $\pm w_{23}$ (grey area) gives
the probability of $|x| \leq  w_{23}$.}
\label{fig_tra}
\end{figure}

Given a time $t_1$ of an event of the detector \#1, the
probability that detector \#2 has an event at a time $t_2$
such that $|t_2-t_1| \leq w_{12}$ is
\begin{equation}
P_{12} = 2 ~\frac{w_{12}}{T_{obs}} N_2
\label{eq_12}
\end{equation}
and similarly for detector \#3.
These two probabilities are independent, so that the probability
that both occur is $P_{12} \cdot P_{13}$.
When both occurs (necessary condition for a triple coincidence)
we can find the distribution of the variable $x = t_3 - t_2$
by considering that the variables $t_2,t_3$ have uniform distributions
in the intervals $\pm w_{12}, \pm w_{13}$ respectively. 
Their probability density functions (p.d.f.) are then
\begin{equation}
F(t_j) = \frac{1}{2 w_{1j}}
\label{eq_pdf}
\end{equation}
with j=2 or 3, and their characteristic functions (Fourier transform
of the p.d.f.) are
\begin{equation}
\Phi_{t_j}(\omega) = \frac{sin(\omega ~w_{1j})}{\omega ~w_{1j}}
\label{eq_chf}
\end{equation}
The characteristic function of the variable
$x = t_3 - t_2$ is
\begin{equation}
\Phi_x(\omega) = \Phi_{t_2}(\omega) \Phi_{t_3}(\omega)
\label{eq_chf2}
\end{equation}
and its p.d.f. F(x) is then given by the inverse Fourier transform of
$\Phi_x(\omega)$
\begin{equation}
F(x) = \frac{1}{2 \pi} \int_{-\infty}^{+\infty} e^{-i \omega x} \Phi_x(\omega) d\omega
\label{eq_invft}
\end{equation}
yielding the trapezoidal shape shown in Fig.\ref{fig_tra} and described by
\begin{eqnarray}
F(x) = \frac{|w_{12} + w_{13} + x| + |w_{12} + w_{13} - x|}{8 w_{12} w_{13}} - \nonumber \\
- \frac{|w_{12} - w_{13} + x| + |w_{12} - w_{13} - x|}{8 w_{12} w_{13}}
\label{eq_f23}
\end{eqnarray}
where we have assumed $w_{12} \geq w_{13}$.

The Probability $P_{23}$ that also detectors \#2 and \#3 are in coincidence, i.e. $|t_2-t_3| \leq w_{23}$, is a fraction of the area of this trapezium: 
\begin{eqnarray}
P_{23} =\frac{w_{23}}{w_{12}},  ~~~~~~~~~~~~~for ~w_{23} \leq (w_{12}-w_{13})\nonumber\\
P_{23} = 1, ~~~~~~~~~~~~~~~~~~for ~w_{23} \geq (w_{12}+w_{13}) \nonumber
\end{eqnarray}
and in the intermediate range $w_{12}-w_{13} < w_{23} < w_{12}+w_{13}$
\begin{eqnarray}
P_{23} = \frac{2(w_{12}w_{13}+w_{12}w_{23}+w_{13}w_{23})-(w_{12}^2+w_{13}^2+w_{23}^2)}{4 w_{12}w_{13}},\nonumber \\
\label{eq_area}
\end{eqnarray}


The probability of a triple coincidence at each event of the detector \#1 is given by the 
product $P_{12} \cdot P_{13} \cdot P_{23}$. The number of accidental triple coincidences
is obtained by further multiplying by $N_1$
\begin{equation}
N_{acc} = 4 P_{23} \frac{w_{12} w_{13} }{T_{o et~al.bs}^2} N_1 N_2 N_3
\label{eq_acc}
\end{equation}
which turns to eq.\ref{eq_3std} when all coincidence windows are equal to $w$.

In our case the coincidence window is set from the timing uncertainties $\sigma_{1,2,3}$
of the single events, according to eq.\ref{eq:coinc}. Then, it is easy to verify
that $w_{ij}$ is bounded between the difference and the sum of the other two
$w$'s so that $P_{23}$ is given by eq.\ref{eq_area}. The resulting analytical estimate for the accidental coincidences in case of constant event rates and different time windows is
\begin{widetext}
\begin{equation}
N_{acc} = \frac{1}{T_{obs}^2} N_1 N_2 N_3 \lbrace2(w_{12}w_{13}+w_{12}w_{23}+w_{13}w_{23})-
(w_{12}^2+w_{13}^2+w_{23}^2)\rbrace
\label{eq_nacc}
\end{equation}
\end{widetext}

\subsection{Implementation}
The common observation time of the three detectors has been 
divided in short sub-intervals with a 
duration randomly chosen within a selected range, e.g. from $\simeq 1/2 hour$ to $\simeq 1 hour$. 
The minimum and maximum duration must be chosen to meet
the assumptions required by eq.A10. In particular, the event rate should be stationary, the coincidence window much smaller than the average distance between events and the number of accidental (background) events much larger than the number of signal (foreground) events.



Since the AURIGA events had a variable time
uncertainty, we computed eq.\ref{eq_nacc} for each
of them using different time windows. The prediction of $N_{acc}$ is obtained by summing the result over all the AURIGA events in the $j^{th}$
sub-interval
\begin{widetext}
\begin{equation}
N_{acc}(j) = \frac{1}{T_j^2} N_{Ex}(j) N_{Na}(j) \sum_{k=1}^{N_{Au}(j)}
 F\left(w(E,N),w(A_k,E),w(A_k,N)\right)
\label{eq_sum}
\end{equation}
\end{widetext}
where $T_j$ is the interval duration, $N_{Ex/Na/Au}$ are the number
of events of Ex, Na, Au, $w(E,N)$ is the Ex-Na (fixed) coincidence
window, $w(A_k,E/N)$ is the coincidence window Au-(Ex or Na)
computed with the $\sigma_t$ of the $k^{th}$ Auriga event, and
$F(w,w,w)$ is the combination of windows in eq.\ref{eq_nacc}.
The total result for the whole overlapping period is then obtained
summing over all the sub-intervals.

This procedure can be repeated with a different choice of the
minimum and maximum intervals duration and/or with a different
random initialization, in order to evaluate the fluctuations 
in the numerical value of the final result.



\end{document}